\documentclass[prd,showpacs,preprintnumbers,eqsecnum,amsmath,amssymb]{revtex4}

\usepackage{graphicx}
\usepackage{dcolumn}
\usepackage{bm}

\newcommand{\be}{\begin{equation}}
\newcommand{\ee}{\end{equation}}
\newcommand{\ba}{\begin{eqnarray}}
\newcommand{\ea}{\end{eqnarray}}

\begin{document}
\title{Dirac quasinormal modes of the Reissner-Nordstr\"om de Sitter black hole}

 \author{Jiliang Jing} \email{jljing@hunnu.edu.cn}
\affiliation{ Institute of Physics and  Department of Physics, \\
Hunan Normal University,
\\ Changsha, Hunan 410081, P. R. China \\
and  \\
School of Mathematics and Statistics,\\ University of Newcastle Upon
Tyne,\\ Newcastle Upon Tyne NE1 7RU, UK}

\vspace*{0.2cm}
\begin{abstract}
\vspace*{0.2cm}

The quasinormal modes of the Reissner-Nordstr\"om de Sitter black
hole for the massless Dirac fields are studied using the
P\"oshl-Teller potential approximation.  We find that the
magnitude of the imaginary part of the quasinormal frequencies
decreases as the cosmological constant or the orbital angular
momentum increases, but it increases as the charge or the overtone
number increases. An interesting feature is that the imaginary
part is almost linearly related to the real part as the
cosmological constant changes for fixed charge, and the linearity
becomes better as the orbital angular momentum increases. We also
prove exactly that the Dirac quasinormal frequencies are the same
for opposite chirality.

\end{abstract}

\vspace*{0.4cm}

\maketitle

\section{INTRODUCTION}
\label{sec:intro} \vspace*{0.2cm}

It is well known that dynamical evolution of field perturbation
on a black hole background can be roughly divided into three
stages \cite{Frolov98}. The first one is an initial wave burst
coming directly from the source of perturbation and is dependent
on the initial form of the original field perturbation. The
second one involves the damped oscillations which frequencies and
damping times are entirely fixed by the structure of the
background spacetime and are independent of the initial
perturbation. This stage can be accurately described in terms of
the discrete set of quasinormal modes (QNMs). And the last one is
a power-law tail behavior of the waves at very late time which is
caused by backscattering of the gravitational field.

QNMs of a black hole are defined as proper solutions of the
perturbation equations belonging to certain complex characteristic
frequencies which satisfy the boundary conditions appropriate for
purely ingoing waves at the event horizon and purely outgoing
waves at infinity \cite{Chand75}. It is generally believed that
QNMs carry a unique footprint to directly identify the black hole
existence. Through the QNMs, one can extract information of the
physical parameters of the black hole---mass, electric charge,
and angular momentum---from the gravitational wave signal by
fitting the few lowest observed quasinormal frequencies to those
calculated from the perturbation analysis, as well as test the
stability of the event horizon against small perturbations.
Moreover, QNMs may be related to fundamental physics, such as the
thermodynamic properties of black holes in loop quantum gravity
\cite{Hod} \cite{Dreyer}.

The idea of QNMs in asymptotically flat black hole started with
the work of Vishveshwara \cite{Vishveshwara}, and the actually
numerical calculation of the quasinormal frequencies was first
presented by Chandrasekhar and Detweiler \cite{Chand75}. Since
then great effort has been contributed to calculate QNMs of
asymptotically flat black holes \cite{Kokkotas} \cite{Nollert}.
QNMs in asymptotically flat spacetimes have recently acquired a
further importance since the real part of the quasinormal
frequencies with a large imaginary part is equal to the
Barbero-Immirzi parameter
\cite{Hod}\cite{Dreyer}\cite{Baez}\cite{Kunstatter}, a factor
introduced by hand in order that loop quantum gravity reproduces
correctly entropy of the black hole.

Recently a new application of the quasinormal mode spectrum has
also arisen from superstring theory \cite{Maldacena} \cite{Witten}
\cite{Kalyana}. There is a suggestion, known as the AdS/CFT
correspondence, that string theory in anti-de Sitter (AdS) space
is equivalent to conformal field theory (CFT) in one less
dimension \cite{Maldacena} \cite{Witten} \cite{Kalyana}.
According to the AdS/CFT correspondence, a large static black
hole in asymptotically AdS spacetime corresponds to a thermal
state in CFT, and the decay of the test field in the black hole
spacetime corresponds to the decay of the perturbed state in CFT.
Thus the quasinormal frequencies give us the thermalization time
scale which is very hard to compute directly. Therefore, many
authors have delved into the studies of QNMs for different
asymptotically AdS black holes \cite{Chan,Birmingham,Cardoso,
Konoplya, Starinets,Kurita,Horowitz, Cardoso1, Cardoso2,
Konoplya1}.

There is considerable observational evidence that the physical
universe has a positive cosmological constant \cite{Perlmutter,
Caldwell, Garnavich}. This observation is at least partially
responsible for the recent flurry of investigations into
asymptotically de Sitter spacetimes. In particular, the study of
QNMs in asymptotically de Sitter spacetimes has garnered much
attention \cite{Moss, Cardoso3, Molina, Maassen, Suneeta,
Zhidenko}. Moss and collaborators \cite{Moss} first carried out
the calculation of the quasinormal frequencies for the
gravitational perturbations of the Schwarzschild de Sitter black
hole. Cardoso and Lemos \cite{Cardoso3} devised an analytical
method to study the case in which the black hole and the
cosmological horizons are very close to each other. Molina
\cite{Molina} extended the analytical method to higher
dimensional Schwarzschild de Sitter black holes, and  Maassen
\cite{Maassen} used it to study the quasinormal mode spectrum of
the Schwarzschild de Sitter black hole for the scalar, the
electromagnetic, and the gravitational fields in the limit of
nearly equal black hole and cosmological radii. Zhidenko
\cite{Zhidenko} calculated the low-laying quasinormal frequencies
of the Schwarzschild de Sitter black hole for fields of different
spin by using the sixth-order WKB and the P\"oshl-Teller
potential approximations. However, the question how the charge of
a asymptotically de Sitter black hole affects the quasinormal
frequencies still remains open. The main purpose of this paper is
to study the question by calculating the Dirac quasinormal
frequencies of the Reissner-Nordstr\"om de Sitter black hole.

The  paper is organized  as follows: In Sec. II, the massless
Dirac field equation in the Reissner-Nordstr\"om  de Sitter black
hole spacetime is decoupled by introducing a tetrad. In Sec. III,
the relation between the quasinormal modes spectra for the Dirac
particles and the antiparticles is discussed. In Sec. IV, the
quasinormal frequencies are calculated and the results are
presented by tables and figures. The last section devotes to
summary and discussions.

\section{Dirac equation in Reissner-Nordstr\"om  de Sitter Black
hole spacetime}
 \vspace*{0.2cm}

In standard coordinates, the metric for the Reissner-Nordstr\"om
de Sitter black hole can be expressed as
\begin{eqnarray}    ds^2=-f dt^2+\frac{1}{f}dr^2+r^2(d\theta^2
+sin^2\theta d\varphi^2),
\end{eqnarray}
with
\begin{eqnarray}
    f=1-\frac{2M}{r}+\frac{Q^2}{r^2}-\frac{\Lambda}{3}r^2,
    \label{f}
\end{eqnarray}
where the parameters $M$, $Q$, and $\Lambda$ represent the black
hole  mass, the charge, and the  cosmological constant, respectively.

 In order to separate the massless Dirac equation \cite{Brill}
 \begin{eqnarray}
  [\gamma^a e_a^\mu(\partial _\mu+\Gamma_\mu)]\Psi=0, \label{Di}
\end{eqnarray}
where $\gamma^a$ is the Dirac matrix,  $e_a^\mu$ is the inverse of
the tetrad $e_\mu^a$, and $\Gamma_\mu $ is the spin connection
which is defined as $\Gamma_\mu= \frac{1}{8}[\gamma^a,\gamma^b]
e_a^\nu e_{b\nu;\mu}$, we take the tetrad as
\begin{eqnarray}
    e_\mu^a=diag(\sqrt{f}, \frac{1}{\sqrt{f}}, r, r \sin \theta).
\end{eqnarray}
Then, the Dirac equation (\ref{Di}) becomes
\begin{eqnarray}
    -\frac{\gamma_0}{\sqrt{f}}\frac{\partial \Psi}{\partial t}+\sqrt{f}
    \gamma_1 \left(\frac{\partial }{\partial r}+\frac{1}{r}+\frac{1}{4 f}
    \frac{d f}{d r} \right) \Psi+\frac{\gamma_2}{r}(\frac{\partial }
    {\partial \theta}+\frac{1}{2}cot\theta)\Psi+\frac{\gamma_3}{r
    \sin\theta}\frac{\partial \Psi}{\partial \varphi}=0. \label{Di1}
\end{eqnarray}
If we define
\begin{eqnarray}
    \Psi=f^{-\frac{1}{4}}\Phi,
\end{eqnarray}
Eq. (\ref{Di1}) can be simplified as
\begin{eqnarray}
    -\frac{\gamma_0}{\sqrt{f}}\frac{\partial \Phi}{\partial t}+\sqrt{f}
    \gamma_1 \left(\frac{\partial }{\partial r}+\frac{1}{r} \right)
    \Phi+\frac{\gamma_2}{r}(\frac{\partial }{\partial \theta}+
    \frac{1}{2}cot\theta)\Phi+\frac{\gamma_3}{r \sin\theta}
    \frac{\partial \Phi}{\partial \varphi}=0. \label{Di2}
\end{eqnarray}
Introducing a tortoise coordinate change
\begin{eqnarray}
    r_*=\int \frac{d r}{f}, \label{tor}
\end{eqnarray}
and the ansatz
\begin{eqnarray}
    \Phi=\left(
\begin{array}{c}
\frac{i G^{(\pm)}(r)}{r}\phi^{\pm}_{jm}(\theta, \varphi) \\
\frac{F^{(\pm)}(r)}{r}\phi^{\mp}_{jm}(\theta, \varphi)
\end{array}\right)e^{-i \omega t},
\end{eqnarray}
with
\begin{eqnarray}
    \phi^{+}_{jm}=\left(
\begin{array}{c}
\sqrt{\frac{j+m}{2 j}}Y^{m-1/2}_l \\
\sqrt{\frac{j-m}{2 j}}Y^{m+1/2}_l
\end{array}\right), \ \ \ \ \ \ \ \ \ \ \ \  (for \ \ j=l+\frac{1}{2}),
\nonumber
\end{eqnarray}
\begin{eqnarray}
    \phi^{-}_{jm}=\left(
\begin{array}{c}
\sqrt{\frac{j+1-m}{2 j+2}}Y^{m-1/2}_l \\
-\sqrt{\frac{j+1+m}{2 j+2}}Y^{m+1/2}_l
\end{array}\right), \ \ \ \ \ \ (for \ \ j=l-\frac{1}{2}), \nonumber
\end{eqnarray}
we find that the cases for $(+)$ and $(-)$ in the functions
$F^{\pm}$ and $G^{\pm}$ can be put together, and then the
decoupled equations can be written as
\begin{eqnarray}
    \frac{d^2 F}{d r_*^2}+(\omega^2-V_1)F&=&0, \label{even}\\
    \frac{d^2 G}{d r_*^2}+(\omega^2-V_2)G&=&0, \label{odd}
\end{eqnarray}
with
\begin{eqnarray}
V_1&=&\frac{\sqrt{f}|k|}{r^2}\left(|k|\sqrt{f}+\frac{r}{2}\frac{d f}
{d r}-f\right), \ \ \ \left(for \ \  k=j+\frac{1}{2},\ \ \ \ \ \  \
\ \ \  and \ \ j=l+\frac{1}{2}\right), \label{V1} \\
V_2&=&\frac{\sqrt{f}|k|}{r^2}\left(|k|\sqrt{f}-\frac{r}{2}\frac{d f}
{d r}+f\right), \ \ \ \left(for \ \  k=-\left(j+\frac{1}{2}\right),\
\  and \ \ j=l-\frac{1}{2}\right). \label{V2}
\end{eqnarray}
The potentials possess two special properties: a) They are
related to the metric function $\sqrt{f}$, which complicate the
calculation of the quasinormal frequencies. b) There is no
quasinormal modes for $l=0$ for potential $V_2$ because $|k|=|l|$
in this case.  If we set $\Lambda=0$ and $Q=0$ the equations
(\ref{even})-(\ref{V2}) give the results of the Schwarzschild
black hole \cite{Cho}.  In the following we will use the
equations (\ref{even})-(\ref{V2}) to study Dirac quasinormal
frequencies.

\section{Relation between the characteristic frequencies
for Dirac particles and antiparticles}
 \vspace*{0.2cm}

In this section we shall prove that Dirac particles and
antiparticles have the same quasinormal mode spectra in the
Reissner-Nordstr\"om de Sitter black hole spacetime.

Introducing a function
\begin{eqnarray}
W=\frac{|k|\sqrt{f}}{r},
\end{eqnarray}
the potentials (\ref{V1}) and (\ref{V2}) can be rewritten as
\begin{eqnarray}
V_1&=&\frac{d W}{dr_*}+W^2, \label{V11}\\
V_2&=&-\frac{d W}{dr_*}+W^2, \label{V22}
\end{eqnarray}
which show that the two potentials are supersymmetric partners
derived from the same superpotential $W$. Now we rewrite metric
function (\ref{f}) as
\begin{eqnarray}
f=\frac{\Lambda}{3r^2}(r-r_-)(r-r_+)(r_c-r)(r-r_b),
\end{eqnarray}
where $r_c$, $r_+$, and $r_-$ represent the cosmological horizon,
the outer event horizon, and the inner event horizon,
respectively, and $r_b$ can be determined by using $r_c$, $r_+$,
and $r_-$. Then the tortoise coordinate (\ref{tor}) can expressed
as
\begin{eqnarray}
r_*&=&\int \frac{d r}{f}\nonumber \\
   &=&\frac{1}{2}\left[\frac{1}{\kappa_-}\ln (r-r_-)
   +\frac{1}{\kappa_+}\ln (r-r_+)-\frac{1}{\kappa_c}\ln (r_c-r)
   +\frac{1}{\kappa_b}\ln (r-r_b)\right], \label{tr}
\end{eqnarray}
where $\kappa_i=\left.\frac{1}{2}\frac{d f}{d r}\right|_{r=r_i}$.
From Eq. (\ref{tr}) we find
\begin{eqnarray}
&& r_*\rightarrow -\infty \ \ \ \ \ \ \    as \ \ \ \ r\rightarrow
r_+, \nonumber \\
&& r_*\rightarrow +\infty \  \ \ \ \ \ \    as \ \ \ \
r\rightarrow r_c.
\end{eqnarray}
Therefore, it is obviously that $W$ is an arbitrary smooth
function over the range of $r_*$, $(-\infty, +\infty)$. And the
superpotential function, together with its derivatives of all
orders with respect to $r_*$, vanish for both $r_*\rightarrow
+\infty $ and $r_*\rightarrow -\infty$. From above discussions we
know that the potentials $V_1$ and $V_2$ also are smooth
functions, integrable over the range of $r_*$, $(-\infty,
+\infty)$, and they vanish exponentially as we approach both the
event horizon $r_+$ and the cosmological horizon $r_c$, i. e.,
\begin{eqnarray}
&& V_{1,2}\rightarrow e^{2 \kappa_+ r_*}, \ \ \ \ \
\ as \ \ r_*\rightarrow -\infty \nonumber\\
&& V_{1,2}\rightarrow e^{-2 \kappa_c r_*}, \ \ \ \ \ as \ \
r_*\rightarrow +\infty. \label{asy}
\end{eqnarray}
Eqs. (\ref{even}), (\ref{odd}), and (\ref{asy}) show that the
asymptotic behaviour of the solutions, $F$ and $G$, for
$r_*\rightarrow \pm \infty$, can be described by
\begin{eqnarray}
e^{\pm i \omega r_*}, \ \ \ \ \ (r_*\rightarrow\pm \infty).
\end{eqnarray}

Ref. \cite{Chandrasekhar} pointed out that there is no
restriction to supposing that, given a solution $G$ of equation
(\ref{odd}),
\begin{eqnarray}
F=pG+q\frac{d G}{d r_*},
\end{eqnarray}
is a solution of equation (\ref{even}), where $p$ and $q$ are
certain suitably chosen functions which satisfy the relations
\cite{Chandrasekhar}
\begin{eqnarray}
&& q(V_1-V_2)=2\frac{d p}{d r_*}+\frac{d^2 q}{d r_*^2}, \label{pq}
\\
&& p^2+\left(p\frac{d q}{d r_*}-q\frac{d p}{d
r_*}\right)-q^2(V_2-\omega^2)=const=C^2. \label{qp}
\end{eqnarray}
For the potentials $V_1$ and $V_2$ we can verify that
\begin{eqnarray}
q=1, \ \ \ p=W,
\end{eqnarray}
satisfy Eqs. (\ref{pq}) and (\ref{qp}) with $C^2=\omega ^2$.
Accordingly, the solutions $F$ and $G$ of Eqs. (\ref{even}) and
(\ref{odd}) are related in the manner
\begin{eqnarray}
i\omega F =W G + \frac{d G}{d r_*}=\frac{|k|\sqrt{f}}{r}G +
\frac{d G}{d r_*}, \label{rel}
\end{eqnarray}
where we have chosen a relative normalization of $F$ and $G$ so
that the inverse relation in the same normalization is
\begin{eqnarray}
i\omega G =-W F + \frac{d F}{d r_*}=-\frac{|k|\sqrt{f}}{r}F +
\frac{d F}{d r_*}.
\end{eqnarray}

The quasinormal modes are defined as solutions of Eqs.
(\ref{even}) and (\ref{odd}), belonging to complex characteristic
frequencies and satisfying the boundary conditions
\begin{eqnarray}
F &&\rightarrow A^{(+)}(\omega)e^{-i\omega r_*}, \ \ \ \ \
r_*\rightarrow  +\infty, \nonumber \\
  && \rightarrow\ \ \ \ \ \ \ \ \ \ e^{+i \omega r_*}, \ \ \ \ \
  r_*\rightarrow -\infty, \nonumber \\
  G &&\rightarrow A^{(-)}(\omega)e^{-i\omega r_*}, \ \ \ \ \
r_*\rightarrow +\infty, \nonumber \\
  && \rightarrow\ \ \ \ \ \ \ \ \ \ e^{+i \omega r_*}, \ \ \ \ \
  r_*\rightarrow -\infty.  \label{rel2}
\end{eqnarray}
If $\omega$ is a characteristic frequency and $G$ is the solution
belonging to it, then the solution $F$ derived from $G$ by Eq.
(\ref{rel}) will satisfy the boundary conditions (\ref{rel2}) with
\begin{eqnarray}
A^{(+)}(\omega)=e^{i\pi}A^{(-)}(\omega).
\end{eqnarray}
Thus, the characteristic frequencies are the same for both $F$
and $G$. Physically the result indicates that Dirac particles and
antiparticles have the same quasinormal mode spectra in the
Reissner-Nordstr\"om de Sitter black hole spacetime. We shall
therefore concentrate just on Eq. (\ref{even}) with potential
$V_1$ in evaluating the quasinormal frequencies in the next
section.

\section{Dirac Quasinormal frequencies in  the Reissner-Nordstr\"om
de Sitter black hole spacetime}
 \vspace*{0.2cm}

Moss and Norman \cite{Moss} found that the gravitational QNMs of
the Schwarzschild-de Sitter black hole calculated by the Leaver
approach \cite{Leaver} agree with the results by using the
P\"oshl-Teller potential approximation for low overtone, and
Zhidenko \cite{Zhidenko} showed that the quasinormal mode
frequencies obtained by the P\"oshl-Teller approximation and the
sixth-order WKB approximation are in a very good agreement for the
Schwarzschild de Sitter black hole. The reason why the results
obtained by the P\"oshl-Teller potential approximation are good
agreement with other numerical results is that the potentials of
black holes in  de Sitter space fall off exponentially.
Therefore, for the Reissner-Nordstr\"om de Sitter black hole we
can study the Dirac quasinormal modes by using the P\"oshl-Teller
approximate potential \cite{Poshl} \cite{Mashhoon} \cite{Ferrari}
\begin{eqnarray}
    V_{PT}=\frac{V_0}{cosh^2(r_*/b)}.
\end{eqnarray}
The potential contains two free parameters ($V_0$ and $b$) which
are used to fit the height and the second derivative of the
potential at the maximum. The location of the maximum has to be
found numerically. From the potentials (\ref{V1}) and (\ref{V2})
we find that the position of the  potential peak as
$|k|\rightarrow \infty$ is
\begin{eqnarray}
    && r_0(|k|\rightarrow \infty)\rightarrow \frac{3}{2}M+\frac{3}{2}M
    \sqrt{1-\frac{8Q^2}{9 M^2}},
\end{eqnarray}
and
\begin{eqnarray}
    && r_0(|k|)\leq r_0(|k|\rightarrow
    \infty), \ \ \  \ \ \ \ (for \ \ V1),    \\ && r_0(|k|)\geq r_0
    (|k|\rightarrow \infty),  \ \ \ \ \ \ \ (for \ \ V2).
\end{eqnarray}
If we use the mass $M$ of the black hole as a unit of mass and
length, the value of the $r_0(|k|\rightarrow \infty)$ just relys
on the charge $Q$. It gives $r_0(|k|\rightarrow \infty)=3$ for an
uncharged black hole, which agrees with result for the
Schwarzschild-de Sitter black hole in Refs. \cite{Moss}
\cite{Cho}. We should point out that although $r_0(|k|\rightarrow
\infty)$ does not depend on the cosmological constant $\Lambda$,
$r_0(|k|)$ does in general.

\begin{figure}
\includegraphics{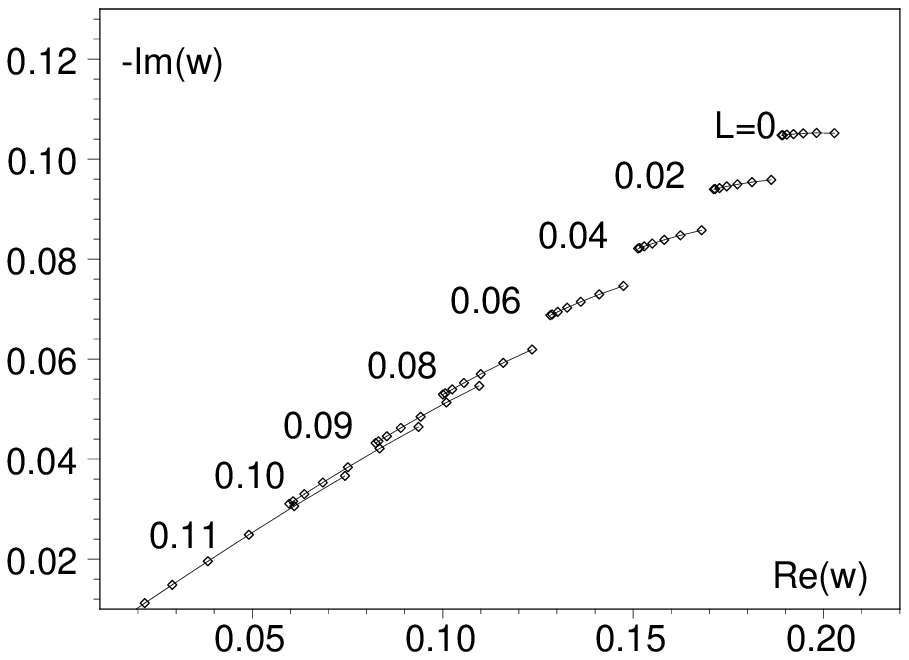}
\caption{\label{PTLK1}Dirac quasinormal mode frequencies for the
potential $V_1$: $|k|=1$, $n=0$. The lines  are drawn  for
$\Lambda=L=0,\ 0.02,\ 0.04, \ 0.06, \ 0.08,\ 0.09,\ 0.10, \ 0.11
$.}
\end{figure}

\begin{figure}
\includegraphics{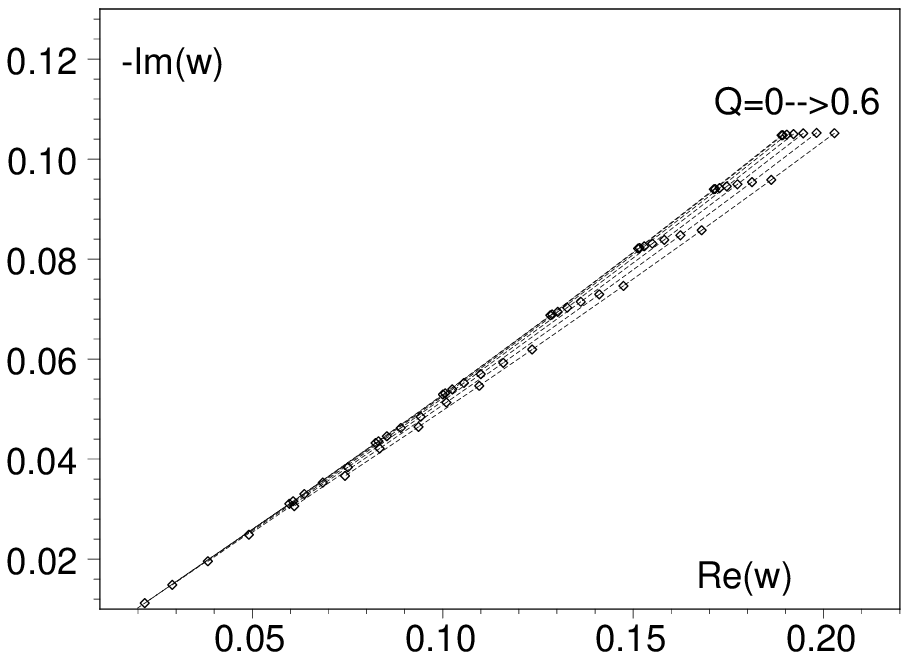}
\caption{\label{PTQK1}Dirac quasinormal mode frequencies for the
potential $V_1$: $|k|=1$, $n=0$. The lines  are drawn for $Q=0, \
0.1, \ 0.2, \ 0.3, \ 0.4, \ 0.5, \ 0.6$.}
\end{figure}

\begin{figure}
\includegraphics{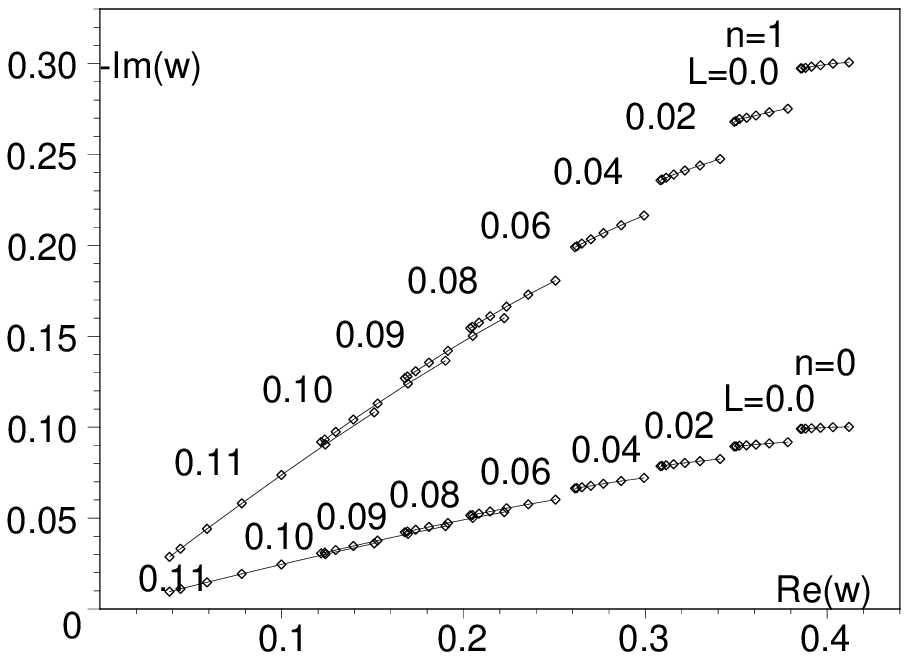}
\caption{\label{PTLK2}Dirac quasinormal mode frequencies for the
potential $V_1$: $|k|=2$. The lines  are drawn  for $\Lambda=L=0,\
0.02,\ 0.04, \ 0.06, \ 0.08,\ 0.09,\ 0.10, \ 0.11 $.}
\end{figure}

\begin{figure}
\includegraphics{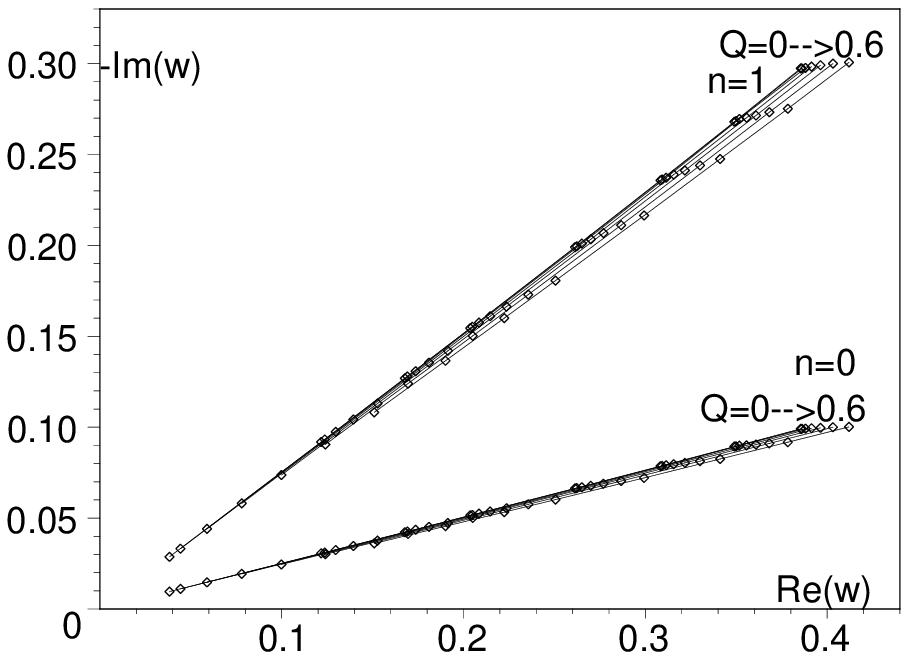}
\caption{\label{PTQK2}Dirac quasinormal mode frequencies for the
potential $V_1$: $|k|=2$.  The lines  are drawn for $Q=0, \ 0.1, \
0.2, \ 0.3, \ 0.4, \ 0.5, \ 0.6$.}
\end{figure}

\begin{figure}
\includegraphics{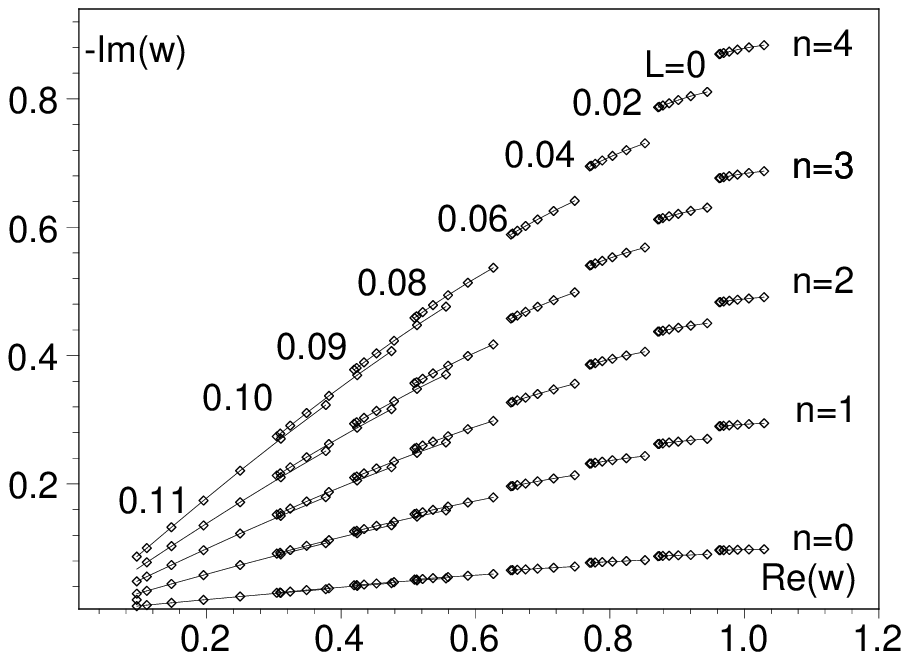}
\caption{\label{PTLK5}Dirac quasinormal mode frequencies for the
potential $V_1$: $|k|=5$. The lines  are drawn  for
$\Lambda=L=0,\ 0.02,\ 0.04, \ 0.06, \ 0.08,\ 0.09,\ 0.10, \ 0.11
$.}
\end{figure}

\begin{figure}
\includegraphics{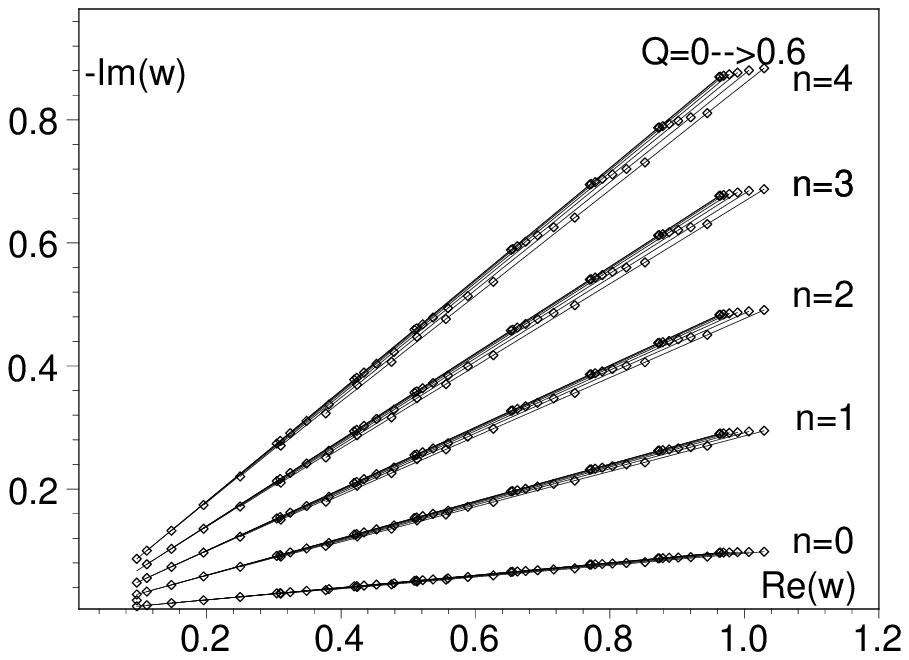}
\caption{\label{PTQK5}Dirac quasinormal mode frequencies for the
potential $V_1$: $|k|=5$.  The lines  are drawn for $Q=0, \ 0.1, \
0.2, \ 0.3, \ 0.4, \ 0.5, \ 0.6$.}
\end{figure}

The quasinormal frequencies of the P\"oschl-Teller potential can
be evaluated analytically \cite{Mashhoon} \cite{Ferrari}
\begin{eqnarray}
    \omega_n=\frac{1}{b}\left[\sqrt{V_0 b^2-\frac{1}{4}}-\left(n+
    \frac{1}{2}\right)i \right], \label{wn} \ \ \ \ \ \ (n=0,
    1,2...).
\end{eqnarray}
However, the quantities $V_0$ and $b$ for different $Q$ and
$\Lambda$ should be found numerically. The quasinormal
frequencies of the Dirac particles in the Reissner-Nordstr\"om de
Sitter black hole spacetime for different $Q$, $\Lambda$, $|k|$,
and $n$ are presented in tables \ref{tab:table1}-\ref{tab:table8}
and figures \ref{PTLK1}-\ref{PTQK5}.

In the calculation we hold 85 digits in all intermediate values
and use the black hole mass units.

\section{Summary and discussions}

The quasinormal mode frequencies of the Reissner-Nordstr\"om de
Sitter black hole for the massless Dirac fields are studied by
using P\"oshl-Teller potential approximation. The results
indicate that the quasinormal frequencies are determined by the
Mass $M$, the charge $Q$, and the cosmological constant $\Lambda$
only. From the tables and figures we find the following
properties for the quasinormal frequencies: a) The magnitude of
the imaginary part of the quasinormal mode frequencies decreases
as the cosmological constant increases for fixed $Q$, $k$, and
$n$, but it increases as the charge increases for fixed
$\Lambda$, $k$, and $n$. b) The magnitude of the imaginary part
decreases as $|k|$ (which relates to the orbital angular
momentum) increases for fixed $n$, $\Lambda$, and $Q$, but it
increases as the overtone number $n$ increases for fixed $|k|$,
$\Lambda$, and $Q$. And c) An interesting feature is that the
imaginary part is almost linearly related to the real part as
cosmological constant changes for fixed charge, and the linearity
becomes better as $|k|$ increases.  We also prove exactly that
the Dirac quasinormal frequencies of the Reissner-Nordstr\"om de
Sitter black hole (with the same $|k|$. We should note that the
orbital angular moment is different, i.e.,  $l_{p}$ for particles
with potential $V_1$ and $l_{ap}=l_{p}+1$ for antiparticles with
potential $V_2$) are the same for opposite chirality.

\begin{acknowledgments}
I would like to thank Prof. Ian G. Moss for helpful discussion.
This work was supported by the National Natural Science Foundation
of China under Grant No. 10275024; and the FANEDD under Grant No.
2003052.
\end{acknowledgments}

\vspace*{2cm}

\begin{table}
\caption{\label{tab:table1} Dirac quasinormal mode frequencies
for potential $V_1$: $|k|=1$, $n=0$}
\begin{ruledtabular}
\begin{tabular}{c|c|c|c|c|c|c|c}
$\omega $ & Q=0.0 & Q=0.1 & Q=0.2 &  Q=0.3 & Q=0.4 & Q=0.5 & Q=0.6 \\
\hline
$\Lambda=0.00$ & 0.1890-0.1048i& 0.1893-0.1048i& 0.1903-0.1049i&
0.1921-0.1050i& 0.1946-0.1051i& 0.1982-0.1052i& 0.2029-0.1052i\\

$\Lambda=0.02$ & 0.1712-0.0939i& 0.1715-0.0940i& 0.1727-0.0942i&
0.1746-0.0945i& 0.1773-0.0949i& 0.1812-0.0954i&  0.1862-0.0959i\\

$\Lambda=0.04$ & 0.1513-0.0821i& 0.1517-0.0822i& 0.1529-0.0825i&
0.1550-0.0831i& 0.1581-0.0838i& 0.1624-0.0847i&  0.1680-0.0858i\\

$\Lambda=0.06$ & 0.1283-0.0688i& 0.1287-0.0689i& 0.1302-0.0695i&
0.1327-0.0703i& 0.1362-0.0715i& 0.1411-0.0729i&  0.1474-0.0746i\\

$\Lambda=0.08$ & 0.1001-0.0529i& 0.1007-0.0532i& 0.1025-0.0539i&
0.1056-0.0552i& 0.1100-0.0570i& 0.1159-0.0593i&  0.1235-0.0619i\\

$\Lambda=0.09$ & 0.0824-0.0432i& 0.0831-0.0436i& 0.0853-0.0446i&
0.0890-0.0462i& 0.0942-0.0485i& 0.1010-0.0513i&  0.1096-0.0547i\\

$\Lambda=0.10$ & 0.0597-0.0311i& 0.0607-0.0315i& 0.0637-0.0330i&
0.0685-0.0353i& 0.0751-0.0384i& 0.0834-0.0421i&  0.0936-0.0464i\\

$\Lambda=0.11$ & 0.0189-0.0097i& 0.0218-0.0112i& 0.0290-0.0148i&
0.0384-0.0196i& 0.0491-0.0249i& 0.0610-0.0306i&  0.0743-0.0366i\\
  \end{tabular}
\end{ruledtabular}
\end{table}

\begin{table}
\caption{\label{tab:table2} Dirac quasinormal mode frequencies
for potential $V_1$: $|k|=2$, $n=0$}
\begin{ruledtabular}
\begin{tabular}{c|c|c|c|c|c|c|c}
$\omega $ & Q=0.0 & Q=0.1 & Q=0.2 &  Q=0.3 & Q=0.4 & Q=0.5 & Q=0.6 \\
\hline
$\Lambda=0.00$ & 0.3855-0.0991i& 0.3861-0.0991i& 0.3881-0.0993i&
0.3915-0.0995i& 0.3964-0.0997i& 0.4031-0.1000i& 0.4121-0.1003i \\

$\Lambda=0.02$ & 0.3490-0.0894i& 0.3497-0.0894i& 0.3519-0.0897i&
0.3556-0.0900i& 0.3610-0.0905i& 0.3684-0.0911i& 0.3783-0.0917i \\

$\Lambda=0.04$ & 0.3083-0.0786i& 0.3091-0.0787i& 0.3115-0.0791i&
0.3157-0.0796i& 0.3218-0.0804i&  0.3301-0.0814i& 0.3411-0.0825i \\

$\Lambda=0.06$ & 0.2612-0.0663i& 0.2622-0.0665i& 0.2651-0.0670i&
0.2700-0.0678i& 0.2771-0.0690i& 0.2867-0.0704i& 0.2993-0.0722i \\

$\Lambda=0.08$ & 0.2037-0.0515i& 0.2049-0.0518i& 0.2086-0.0525i&
0.2148-0.0537i& 0.2237-0.0555i& 0.2355-0.0576i& 0.2506-0.0603i \\

$\Lambda=0.09$ & 0.1677-0.0423i& 0.1692-0.0426i& 0.1737-0.0436i&
0.1811-0.0452i& 0.1915-0.0474i& 0.2051-0.0501i& 0.2224-0.0534i \\

$\Lambda=0.10$ & 0.1217-0.0306i& 0.1237-0.0311i& 0.1297-0.0325i&
0.1394-0.0347i& 0.1527-0.0377i& 0.1695-0.0413i& 0.1900-0.0455i \\

$\Lambda=0.11$ & 0.0384-0.0096i& 0.0444-0.0111i& 0.0590-0.0147i&
0.0781-0.0194i& 0.0999-0.0246i& 0.1241-0.0302i& 0.1509-0.0361i \\
  \end{tabular}
\end{ruledtabular}
\end{table}

 \begin{table}
\caption{\label{tab:table3} Dirac quasinormal mode frequencies
for potential $V_1$: $|k|=2$, $n=1$}
\begin{ruledtabular}
\begin{tabular}{c|c|c|c|c|c|c|c}
$\omega $ & Q=0.0 & Q=0.1 & Q=0.2 &  Q=0.3 & Q=0.4 & Q=0.5 & Q=0.6 \\
\hline
$\Lambda=0.00$ & 0.3855-0.2972i& 0.3862-0.2974i& 0.3881-0.2978i&
0.3915-0.2984i& 0.3964-0.2992i& 0.4031-0.3001i& 0.4121-0.3008i\\

$\Lambda=0.02$ & 0.3490-0.2681i& 0.3497-0.2683i& 0.3519-0.2689i&
0.3556-0.2700i& 0.3610-0.2715i& 0.3684-0.2733i& 0.3783-0.2753i\\

$\Lambda=0.04$ & 0.3083-0.2358i& 0.3091-0.2361i& 0.3115-0.2372i&
0.3157-0.2388i& 0.3218-0.2412i&  0.3301-0.2441i& 0.3411-0.2475i\\

$\Lambda=0.06$ & 0.2612-0.1990i& 0.2622-0.1995i& 0.2651-0.2010i&
0.2700-0.2035i& 0.2771-0.2069i& 0.2867-0.2113i& 0.2993-0.2165i\\

$\Lambda=0.08$ & 0.2037-0.1545i& 0.2049-0.1552i& 0.2086-0.1575i&
0.2148-0.1612i& 0.2237-0.1664i& 0.2355-0.1729i& 0.2506-0.1807i\\

$\Lambda=0.09$ & 0.1677-0.1269i& 0.1692-0.1279i& 0.1737-0.1308i&
0.1811-0.1356i& 0.1915-0.1421i& 0.2051-0.1504i& 0.2224-0.1601i\\

$\Lambda=0.10$ & 0.1217-0.0918i& 0.1237-0.0932i& 0.1297-0.0974i&
0.1394-0.1041i& 0.1527-0.1131i& 0.1695-0.1240i& 0.1900-0.1366i\\

$\Lambda=0.11$ & 0.0384-0.0289i& 0.0444-0.0334i& 0.0590-0.0442i&
0.0781-0.0582i& 0.0999-0.0738i& 0.1241-0.0906i& 0.1509-0.1082i\\
  \end{tabular}
\end{ruledtabular}
\end{table}

\begin{table}
\caption{\label{tab:table4} Dirac quasinormal mode frequencies for
potential $V_1$: $|k|=5$, $n=0$}
\begin{ruledtabular}
\begin{tabular}{c|c|c|c|c|c|c|c}
$\omega $ & Q=0.0 & Q=0.1 & Q=0.2 &  Q=0.3 & Q=0.4 & Q=0.5 & Q=0.6 \\
\hline
$\Lambda=0.00$ & 0.9625-0.0966i& 0.9641-0.0967i& 0.9690-0.0968i&
0.9774-0.0971i& 0.9898-0.0974i& 1.0067-0.0978i& 1.0293-0.0982i \\

$\Lambda=0.02$ & 0.8719-0.0874i& 0.8734-0.0875i& 0.8788-0.0878i&
0.8880-0.0882i& 0.9016-0.0887i& 0.9202-0.0894i& 0.9448-0.0901i\\

$\Lambda=0.04$ & 0.7700-0.0772i& 0.7720-0.0773i& 0.7781-0.0777i&
0.7885-0.0782i& 0.8038-0.0790i& 0.8246-0.0800i& 0.8520-0.0812i\\

$\Lambda=0.06$ & 0.6528-0.0654i& 0.6551-0.0656i& 0.6623-0.0661i&
0.6746-0.0669i& 0.6923-0.0680i& 0.7163-0.0695i& 0.7478-0.0712i\\

$\Lambda=0.08$ & 0.5092-0.0510i& 0.5123-0.0512i& 0.5214-0.0520i&
0.5369-0.0532i& 0.5591-0.0549i& 0.5885-0.0571i& 0.6265-0.0596i\\

$\Lambda=0.09$ & 0.4195-0.0420i& 0.4232-0.0423i& 0.4342-0.0433i&
0.4527-0.0448i& 0.4787-0.0470i& 0.5129-0.0497i& 0.5559-0.05290i\\

$\Lambda=0.10$ & 0.3043-0.0305i& 0.3096-0.0309i& 0.3243-0.0323i&
0.3486-0.0345i& 0.3819-0.0375i& 0.4239-0.0411i& 0.4751-0.0452i\\

$\Lambda=0.11$ & 0.0962-0.0096i& 0.1112-0.0111i& 0.1477-0.0147i&
0.1955-0.0194i& 0.2500-0.0245i& 0.3103-0.0301i& 0.3773-0.0359i\\
  \end{tabular}
\end{ruledtabular}
\end{table}

\begin{table}
\caption{\label{tab:table5} Dirac quasinormal mode frequencies for
potential $V_1$: $|k|=5$, $n=1$}
\begin{ruledtabular}
\begin{tabular}{c|c|c|c|c|c|c|c}
$\omega $ & Q=0.0 & Q=0.1 & Q=0.2 &  Q=0.3 & Q=0.4 & Q=0.5 & Q=0.6 \\
\hline
$\Lambda=0.00$ & 0.9625-0.2899i& 0.9641-0.2901i& 0.9690-0.2905i&
0.9774-0.2913i& 0.9898-0.2923i& 1.0067-0.2934i& 1.0293-0.2946i\\

$\Lambda=0.02$ & 0.8716-0.2623i& 0.8734-0.2626i& 0.8788-0.2633i&
0.8880-0.2645i& 0.9016-0.2661i& 0.9202-0.2681i& 0.9448-0.2703i\\

$\Lambda=0.04$ & 0.7700-0.2316i& 0.7720-0.2319i& 0.7781-0.2330i&
0.7885-0.2347i& 0.8038-0.2371i& 0.8246-0.2401i& 0.8520-0.2436i\\

$\Lambda=0.06$ & 0.6528-0.1962i& 0.6551-0.1967i& 0.6623-0.1982i&
0.6746-0.2006i& 0.6923-0.2041i& 0.7163-0.2085i& 0.7478-0.2137i\\

$\Lambda=0.08$ & 0.5092-0.1530i& 0.5123-0.1537i& 0.5214-0.1559i&
0.5369-0.1596i& 0.5591-0.1647i& 0.5885-0.1712i& 0.6265-0.1789i\\

$\Lambda=0.09$ & 0.4195-0.1260i& 0.4232-0.1269i& 0.4342-0.1298i&
0.4527-0.1345i& 0.4787-0.1410i& 0.5129-0.1491i& 0.5559-0.1587i\\

$\Lambda=0.10$ & 0.3043-0.0913i& 0.3094-0.0928i& 0.3243-0.0969i&
0.3486-0.1036i& 0.3819-0.1124i& 0.4239-0.1232i& 0.4751-0.1356i\\

$\Lambda=0.11$ & 0.0962-0.0289i& 0.1112-0.0333i& 0.1477-0.0441i&
0.1955-0.0580i& 0.2500-0.0736i& 0.3103-0.0902i& 0.3773-0.1077i\\
  \end{tabular}
\end{ruledtabular}
\end{table}

\begin{table}
\caption{\label{tab:table6} Dirac quasinormal mode frequencies
for potential $V_1$: $|k|=5$, $n=2$}
\begin{ruledtabular}
\begin{tabular}{c|c|c|c|c|c|c|c}
$\omega $ & Q=0.0 & Q=0.1 & Q=0.2 &  Q=0.3 & Q=0.4 & Q=0.5 & Q=0.6 \\
\hline
$\Lambda=0.00$ & 0.9625-0.4832i& 0.9641-0.4834i& 0.9689-0.4842i&
0.9774-0.4855i& 0.9898-0.4871i& 1.0067-0.4891i& 1.0293-0.4910i\\

$\Lambda=0.02$ & 0.8716-0.4372i& 0.8734-0.4376i& 0.8788-0.4388i&
0.8880-0.4408i& 0.9016-0.4435i& 0.9202-0.4468i& 0.9448-0.4505i\\

$\Lambda=0.04$ & 0.7700-0.3860i& 0.7720-0.3866i& 0.7781-0.3883i&
0.7885-0.3912i& 0.8038-0.3951i& 0.8246-0.4001i& 0.8520-0.4060i\\

$\Lambda=0.06$ & 0.6528-0.3270i& 0.6551-0.3278i& 0.6623-0.3303i&
0.6746-0.3344i& 0.6923-0.3401i& 0.7163-0.3474i& 0.7478-0.3562i\\

$\Lambda=0.08$ & 0.5092-0.2549i& 0.5123-0.2562i& 0.5214-0.2599i&
0.5369-0.2660i& 0.5591-0.2745i& 0.5885-0.2853i& 0.6265-0.2982i\\

$\Lambda=0.09$ & 0.4194-0.2099i& 0.4232-0.2115i& 0.4342-0.2163i&
0.4527-0.2242i& 0.4787-0.2350i& 0.5129-0.2485i& 0.5559-0.2646i\\

$\Lambda=0.10$ & 0.3043-0.1522i& 0.3094-0.1546i& 0.3243-0.1615i&
0.3486-0.1726i& 0.3819-0.1874i& 0.4239-0.2053i& 0.4751-0.2260i\\

$\Lambda=0.11$ & 0.0962-0.0481i& 0.1112-0.0555i& 0.1477-0.0736i&
0.1955-0.0967i& 0.2500-0.1226i& 0.3103-0.1503i& 0.3773-0.1795i\\
  \end{tabular}
\end{ruledtabular}
\end{table}

\begin{table}
\caption{\label{tab:table7} Dirac quasinormal mode frequencies
for potential $V_1$: $|k|=5$, $n=3$}
\begin{ruledtabular}
\begin{tabular}{c|c|c|c|c|c|c|c}
$\omega $ & Q=0.0 & Q=0.1 & Q=0.2 &  Q=0.3 & Q=0.4 & Q=0.5 & Q=0.6 \\
\hline
$\Lambda=0.00$ & 0.9625-0.6764i& 0.9641-0.6768i& 0.9689-0.6779i&
0.9774-0.6796i& 0.9898-0.6820i& 1.0067-0.6847i& 1.0293-0.6874i\\

$\Lambda=0.02$ & 0.8716-0.6121i& 0.8734-0.6127i& 0.8788-0.6143i&
0.8880-0.6171i& 0.9016-0.6208i& 0.9202-0.6255i& 0.9448-0.6307i\\

$\Lambda=0.04$ & 0.7700-0.5404i& 0.7720-0.5412i& 0.7781-0.5436i&
0.7885-0.5476i& 0.8038-0.5532i& 0.8246-0.5602i& 0.8520-0.5684i\\

$\Lambda=0.06$ & 0.6528-0.4578i& 0.6551-0.4590i& 0.6623-0.4624i&
0.6746-0.4682i& 0.6923-0.4762i& 0.7163-0.4864i& 0.7478-0.4986i\\

$\Lambda=0.08$ & 0.5092-0.3569i& 0.5123-0.3587i& 0.5214-0.3638i&
0.5369-0.3724i& 0.5591-0.3843i& 0.5885-0.3994i& 0.6265-0.4175i\\

$\Lambda=0.09$ & 0.4195-0.2939i& 0.4232-0.2962i& 0.4342-0.3029i&
0.4527-0.3139i& 0.4787-0.3290i& 0.5129-0.3479i& 0.5559-0.3704i\\

$\Lambda=0.10$ & 0.3043-0.2131i& 0.3094-0.2164i& 0.3243-0.2261i&
0.3486-0.2417i& 0.3819-0.2623i& 0.4239-0.2875i& 0.4751-0.3165i\\

$\Lambda=0.11$ & 0.0962-0.0674i& 0.1112-0.0778i& 0.1477-0.1030i&
0.1955-0.1354i& 0.2500-0.1716i& 0.3103-0.2104i& 0.3773-0.2513i\\
  \end{tabular}
\end{ruledtabular}
\end{table}

\begin{table}
\caption{\label{tab:table8} Dirac quasinormal mode frequencies
for potential $V_1$: $|k|=5$, $n=4$}
\begin{ruledtabular}
\begin{tabular}{c|c|c|c|c|c|c|c}
$\omega $ & Q=0.0 & Q=0.1 & Q=0.2 &  Q=0.3 & Q=0.4 & Q=0.5 & Q=0.6 \\
\hline
$\Lambda=0.00$ & 0.9625-0.8697i& 0.9641-0.8702i& 0.9689-0.8716i&
0.9774-0.8738i& 0.9898-0.8768i& 1.0067-0.8803i& 1.0293-0.8838i\\

$\Lambda=0.02$ & 0.8716-0.7870i& 0.8734-0.7877i& 0.8788-0.7899i&
0.8880-0.7934i& 0.9016-0.7982i& 0.9202-0.8042i& 0.9448-0.8108i\\

$\Lambda=0.04$ & 0.7700-0.6948i& 0.7720-0.6958i& 0.7781-0.6989i&
0.7885-0.7041i& 0.8038-0.7112i& 0.8246-0.7203i& 0.8520-0.7308i\\

$\Lambda=0.06$ & 0.6528-0.5886i& 0.6551-0.5901i& 0.6623-0.5945i&
0.6746-0.6019i& 0.6923-0.6122i& 0.7163-0.6254i& 0.7478-0.6411i\\

$\Lambda=0.08$ & 0.5092-0.4589i& 0.5123-0.4611i& 0.5214-0.4678i&
0.5369-0.4788i& 0.5591-0.4941i& 0.5885-0.5135i& 0.6265-0.5368i\\

$\Lambda=0.09$ & 0.4195-0.3779i& 0.4232-0.3808i& 0.4342-0.3894i&
0.4527-0.4035i& 0.4787-0.4230i& 0.5129-0.4473i& 0.5559-0.4762i\\

$\Lambda=0.10$ & 0.3043-0.2740i& 0.3094-0.2783i& 0.3243-0.2907i&
0.3486-0.3107i& 0.3819-0.3373i& 0.4239-0.3696i& 0.4751-0.4069i\\

$\Lambda=0.11$ & 0.0962-0.0866i& 0.1112-0.1000i& 0.1477-0.1324i&
0.1955-0.1741i& 0.2500-0.2207i& 0.3103-0.2705i& 0.3773-0.3230i\\
  \end{tabular}
\end{ruledtabular}
\end{table}

\end{document}